\documentclass[letterpaper, 11pt]{article}
\usepackage{amssymb,amsmath}
\usepackage{epsf}
\usepackage{times,bm,mathrsfs}  
\usepackage{amsmath}
\usepackage{amssymb}
\usepackage{amsfonts}
\usepackage{mathrsfs}
\usepackage{color}
\usepackage{graphicx}
\usepackage{amscd}
\usepackage{epsfig}

\author{
Christel Kamp\footnote{Paul-Ehrlich-Institut, Federal Institute for Vaccines and Biomedicines, Paul-Ehrlich-Stra{\ss}e 51-59, 63225 Langen, Germany, christel.kamp@pei.de}, Mathieu Moslonka-Lefebvre\footnote{INRA, UR 341 Mathématiques et Informatique Appliquées, 78350 Jouy-en-Josas, France}, Samuel Alizon\footnote{Laboratoire MIVEGEC (UMR CNRS 5290, IRD 224, UM1, UM2), 911 avenue Agropolis, 34394 Montpellier Cedex 5, France}
}

\title{\vspace{0cm}
Predicting epidemics on weighted networks}
\begin{document}

\maketitle
\begin{abstract} 
The contact structure between hosts has a critical influence on disease spread. However, most network-based models  used in epidemiology tend to ignore heterogeneity in the weighting of contacts. This assumption is known to be at odds with the data for many contact networks (e.g.~sexual contact networks) and to have a strong effect on the predictions of epidemiological models. One of the reasons why models usually ignore heterogeneity in transmission is that we currently lack tools to analyze weighted networks, such that most studies rely on numerical simulations. Here, we present a novel framework to estimate key epidemiological variables, such as the rate of early epidemic expansion ($r_0$) and the basic reproductive ratio ($R_0$), from joint probability distributions of number of partners (contacts) and number of interaction events through which contacts are weighted. This framework also allows for a derivation of the full time course of epidemic prevalence and contact behaviour which is validated using numerical simulations.  Our framework allows for the incorporation of more realistic contact networks into epidemiological models, thus improving predictions on the spread of emerging infectious diseases.
\end{abstract}

\section{Introduction}

Contact structure between hosts is known to have a key influence on disease spread \cite{AndersonMay1991}. A striking result is for instance that the more heterogeneous the contact network is, i.e.~the higher the variance in the number of contacts per individual, the more rapid the initial disease spread. 

One way to capture contact structure is to use a network \cite{Newman2002}. Such contact networks are typically described by a square binary adjacency matrix, where each term on the $i$th line and $j$th column can take the value 0 or 1  to indicate respectively the absence or the presence of a contact between individuals $i$ and $j$. The sum of the rows (or the columns) indicates the total number of contacts of an individual (i.e.~the node's degree in the network). Contact networks are widely used because they have several valuable properties, one of which being that the dominant eigenvalue of the adjacency matrix is an indicator of the early propagation of an infectious disease spreading on this network \cite{RestrepoEtal2006,Chakrabarti2008ACM,Moslonkalefebvre2012MB}. 

Insights into epidemic thresholds can be gained through the distribution of the number of contacts (degrees). The number of secondary infections generated by a typical infected host in a fully susceptible population, i.e. the basic reproductive number $R_0$ \cite{AndersonMay1991}, scales with the ratio of the second moment $\langle k^2\rangle$  and first moment (mean) $\langle k\rangle$  of the distribution in the number of contacts $k$. This result holds both for static networks (denoted $R_0^{stat}$)\cite{Durrett2007} as well as for fully mixed, dynamic networks (denoted $R_0^{mix}$) \cite{MayAnderson1987,Moreno2002}. The static case corresponds to networks in which the identity of contacts is fixed (as approximatively seen in sexual contact networks) and the fully mixed dynamic case corresponds to a situation in which individuals update their contacts dynamically in a fully mixed fashion within the population (as approximatively seen in airborne infections on small geographical scales). The number of contacts follows the distribution of $k$ for both network types.  Regarding disease spread, it has been shown that:
\begin{subequations}\label{R0classical}
\begin{eqnarray}
R_0^\text{mix}~\!\!\!\!&=&~\!\!\!\!\frac{\beta}{\gamma}~\frac{\langle k^2\rangle}{\langle k\rangle}=\frac{\beta}{\gamma}~\left(\frac{\sigma_k^2}{\langle k\rangle}+\langle k\rangle \right)\\
R_0^\text{stat}\!\!\!\!&=& \!\!\!\!\frac{\beta}{\beta+\gamma}~\left(\!\frac{\langle k^2\rangle}{\langle k\rangle}-1\!\right) =  \frac{\beta}{\beta+\gamma}~\left(\!\frac{\sigma_k^2}{\langle k\rangle}+\langle k \rangle -1\!\right)
\end{eqnarray}
\end{subequations}
where $\sigma_k^2=\langle k^2\rangle -\langle k\rangle^2$ is the variance of the distribution of the number of contacts.

$R_0^\text{stat}$ and $R_0^\text{mix}$ represent the lower and upper bounds of the basic reproductive ratio \cite{VolzMeyers2009} for SIR epidemics on random networks if individuals transmit the infection at a rate $\beta$ and recover from the infection at a rate $\gamma$.  Findings for both static and dynamic networks imply that heterogeneous networks with a large or even diverging variance in the distribution of the number of contacts have a very small or even vanishing epidemic threshold. 

One of the typical key assumptions epidemiological models on networks make in order to derive such elegant expressions for $R_0$ is that the transmission rate is the same between all pairs of individuals. This is materialized by the fact that all the edges of the contact matrix have the same weight (0 or 1). In reality, however, this is known to be a simplifying assumption \cite{BarratEtal2004b}. Just to take one example related to infectious diseases, in the case of sexual contact networks, the number of sex acts per unit of time is not constant in all partnerships \cite{BlowerBoe1993,NordvikLiljeros2006,BrittonEtal2007}. More generally, the number of interaction events (which correspond to potential transmission events) may vary among contact pairs. Simplifying the reality is  commendable as long as it does not  modify the conclusions of the model. The problem here is that tempering with the weighting of the network  has been shown to lead to the loss of important epidemiological properties of heterogeneous networks, such as the low value of the epidemiological threshold or the negative correlation between network size and the epidemiological threshold value  \cite{Moslonka-LefebvreEtal2012}. 

 An increasing number of studies in epidemiology point to the importance of considering weighted networks. Some examples include the spread of sexually-transmitted infections \cite{Moslonka-LefebvreEtal2012}, disease transmission in sheep flocks \cite{SchleyEtal2012}, respiratory diseases of humans \cite{StehleEtal2011} or general infectious diseases of human  spreading on a social contact network \cite{EamesEtal2009} or on airline connection networks \cite{ColizzaEtal2006}. Several more conceptual studies have also been published in the theoretical physics literature (e.g. \cite{Newman2002,JooLebowitz2004,WangEtal2007}). Most of these studies have in common that they use weighted networks and resort to (heavy) numerical simulations. Indeed, contrary to unweighted networks, we lack analytical frameworks to analyze epidemic spread on weighted networks.

Amongst earlier epidemic models on weighted networks, one approach stands out \cite{BrittonEtal2011}. The authors describe the early epidemic expansion in a Reed-Frost model where infections take place in discrete time steps, with non-overlapping generations and each infected individual recovers  with certainty one time step after infection. These simplifications allow them to  obtain an analytical expression of $R_0$ and to assess outbreak probabilities using branching processes. In their formalism, as shown in \cite{Deijfen2011}, $R_0$ can be derived as the dominant eigenvalue of the mean offspring matrix ($m_{d,k\geq 2}$), where $m_{d,k}$ represents the expected number of individuals with $k$ contacts that an individual with $d$ contacts infects considering potentially degree-dependent network weights. Importantly, it is only because the authors make strong simplifying assumptions  in their model, such as the independence between network weights and nodes' degrees, that they can derive an explicit form of $R_0$.

\begin{figure}[t]
\begin{center}
\includegraphics[angle=0, height = 10cm]{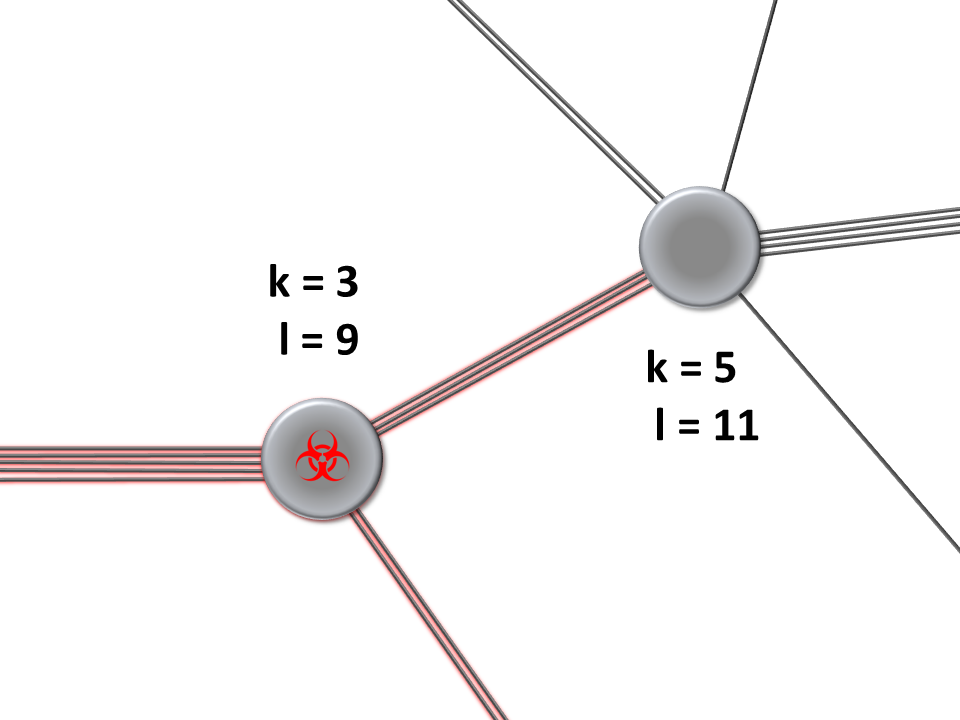}
\caption{Not only is the number of contacts that an individual maintains relevant for the spread of an infectious agent but also the weight that (s)he assigns to each contact. Here, each individual has $l$ interaction events that (s)he can distribute among his/her $k$ contacts. On the scale of the transmission network, these are  modelled by the joint probability distribution $P_{kl}$ to find an individual with $k$ contacts and $l$  interaction events per time interval.\label{fig0}}
\end{center}
\end{figure}

Here, we present an original framework, which builds on configuration type network epidemic approaches \cite{Volz2008,Kamp2010} to model the dynamics of a disease spreading on a weighted network and to estimate key epidemiological variables from the network's properties.  Our framework does not require strong simplifying assumptions regarding the epidemiological process or the distribution of weights on the contact network to provide explicit expressions for $R_0$ or epidemic prevalence.
It goes beyond earlier frameworks describing continuous time SIR epidemics on weighted networks by providing explicit expressions for the rate of early epidemic expansion ($r_0$) and the basic reproductive ratio ($R_0$) of the infection. It also allows for a derivation of the full time course of epidemic prevalence and contact behaviour  of susceptible, infected and recovered individuals (in terms of the probability generating functions (PGFs) of the degree distributions). As sketched in Fig.~\ref{fig0}, the parametrisation is done in a general fashion using the joint probability distribution $P_{kl}$ of an individual to have $k$ contacts among which (s)he randomly distributes $l$  interaction events. We validate our analytical results using numerical simulations.

\section{The Model}

A susceptible individual has $k$ contacts and $l$  interaction events per time interval  (which are distributed among his/her $k$ contacts). This individual can get infected at a rate proportional to the number of interaction events ($l$),  the transmission rate of the pathogen ($\beta$) and the probability for each of his/her contacts to  points to an infected individual ($p_{SI}$). The population dynamics of susceptible, infected and recovered individuals with $k$ contacts and $l$ interaction events per time interval are thus captured by the following set of differential equations:
\begin{subequations} \label{eq:s1}
\begin{eqnarray}
\dot{S}_{kl} &=& -\beta \, p_{SI} \, l \, S_{kl}    \\
\dot{I}_{kl} &=& \beta \, p_{SI} \, l \, S_{kl}   - \gamma \, I_{kl}\\
\dot{R}_{kl} &=& \gamma \, I_{kl},
\end{eqnarray}
\end{subequations}
where $\gamma$ is the rate at which recovery from infection occurs (see Table  \ref{tab:modelnotation}).

\begin{table*}[b!]
\caption{Notations used in the study. \label{tab:modelnotation}}
\begin{tabular*}{\hsize}
{@{\extracolsep{\fill}}lp{9cm}}
\hline\\
$\dot{f}(x,t)=\frac{\partial}{\partial t}f(x,t)$ & partial derivative of function $f$  with respect to $t$\\
$f^{(a,b)}(x,y,t)=\frac{\partial^a}{\partial x^a }\frac{\partial^b}{\partial y^b }f(x,y,t)$ & partial derivative of function $f$ with respect to $x$\\
$A_{kl}$  & number of individuals in group $A$ with $k$ contacts and $l$ (potential)transmission events (per time interval)\\
$A=\sum_{k, l} A_{kl}$ & number of individuals in group $A$\\
$N_{kl}=\sum_A A_{kl}$ & number of individuals with $k$ contacts and $l$ transmission events (per time interval)\\
$N=\sum_{k,l} N_{kl}$ & total number of individuals \\
$P_{Akl}=\frac{A_{kl}}{A}$& probability for an individual in group $A$ to have $k$ contacts and $l$ transmission events per time interval\\
$G_A(x,y,t) = \sum_{k,l} P_{Akl}(t)x^ky^l$ &  probability generating function (PGF) of $P_{Akl}(t)$ \\
$\langle k\rangle_A = G^{(1,0)}_A(1,1,t)$ & average number of contacts of $A$ individuals\\
$\langle l\rangle_A = G^{(0,1)}_A(1,1,t)$ & average number of transmission events per time interval of $A$ individuals\\
$\langle kl\rangle_A = G^{(1,1)}_A(1,1,t)$ & average number of contacts times transmission events per time interval of $A$ individuals\\
$P_{kl}=\frac{N_{kl}}{N}$& probability for an individual to have $k$ contacts and $l$ (potential) transmission events per time interval\\
$G(x,y,t) = \sum_{k,l} P_{kl}(t)x^ky^l=\sum_A \frac{A}{N}g_A(x,t)$ &  probability generating function (PGF) of $p_{kl}(t)$ \\
$\langle k\rangle= G^{(1,0)}(1,1,t)$ & average number of contacts of individuals\\
$M_A=\sum_{k,l} k A_{kl}=AG^{(1,0)}_A(1,1,t)$ & number of links coming from $A$ individuals\\
$M=\sum_A M_A$& number of links\\
$M_{AB}$ & number of links coming from $A$ individuals and pointing to $B$ individuals\\
$p_{AB}=\frac{M_{AB}}{M_A}$ & probability for a link starting from an $A$ individual to point to an $B$ individual\\ 
\hline 
\multicolumn{2}{l}{$A$,$B$ correspond to epidemic stages, i.e. $S$, $I$, $R$ for susceptible, infected, recovered}
\end{tabular*}
\end{table*}

The dynamics of the total number of susceptible, infected and recovered individuals can be obtained by summing over $k$ and $l$. This leads to:
\begin{subequations} \label{eq:s2}
\begin{eqnarray}
\dot{S}&=&-\beta \, p_{SI}   \,  \langle l\rangle_S \, S  \label{S}\\
\dot{I} &=& \beta \, p_{SI}  \, \langle l\rangle_S \, S - \gamma \, I \label{I}\\
\dot{R}&=& \gamma \, I \label{R}.
\end{eqnarray}
\end{subequations}
where $\langle l\rangle_S=\sum_{k,l} l \, S_{kl}/S=G_S^{(0,1)}(1,1,t)$ is the average number of interaction events per time a susceptible individual has.

To close the equation system \ref{eq:s2}, we need expressions for the temporal  dynamics of the $p_{AB}$, i.e.~the probabilities for a status $A$ individual's contact to be with an individual of status $B$. These can be  derived through a careful assessment of the links/contacts among susceptible and infected individuals over the course of an epidemic. This means following the dynamics of the joint probability distribution to find $k$ contacts and $l$ interaction events per time among susceptible individuals $P_{Skl}=S_{kl}/S$ through its PGF $G_S(x,y,t)$.
 The temporal dynamics of the probability generating function $G_S(x,y,t)$  can be calculated by observing that the dynamics of the corresponding joint probability distribution of contacts and  interaction events of susceptible individuals are governed by the equation $\dot{P}_{Skl}=\frac{\dot{S}_{kl}}{S}-\frac{\dot{S}}{S}P_{Skl}$.  As shown, in detail in the Supporting Information, we close equation system \ref{eq:s2} with the following equations:

\begin{subequations} \label{eq:s3}
\begin{eqnarray}
\dot{p}_{SI} \!\!\!\!\! &=& \!\!\!\!\! p_{SI} \,\! \left(\!\beta \, p_{SS}  \,  \frac{\langle kl\rangle_S}{\langle k\rangle_S} - \beta \, (1-p_{SI}-p_{SS})  \, \frac{\langle l\rangle_S}{\langle k\rangle_S}  -\gamma \! \right)\\
\dot{p}_{SS} \!\!\!\!\!&=&\!\!\!\!\! -\beta  \,  p_{SI}  \, p_{SS} \, \frac{\langle kl\rangle_S-2 \langle l\rangle_S}{\langle k\rangle_S}\\
\dot{G}_S(x,\!y,\!t)\!\!\!\!&=&\!\!\!\!\beta  \,  p_{SI}  \, (\langle l\rangle_S \, G_S(x,\!y,\!t)-y \, G_S^{(0,1)}(x,\!y,\!t)).
\end{eqnarray}
\end{subequations}
For furthers details about the terms in this equation system, see Table \ref{tab:modelnotation}.

\section{Validating the analytical model \label{validation}}

 The joint probability distribution of the number of partners and number of interaction events ($P_{kl}$) can be written as the product $P_{kl}=P_k~P_{l|k}$, $P_{l|k}$ being the probability distribution of the number of  interaction events per time $l$ given that the individual has $k$ contacts. If $P_{l|k}=\delta_{lk}$, where $\delta_{lk}$ is 1 if $l=k$ and 0 otherwise, we  are back to a `classical' network case, with an exact linear dependency between the number of contacts and the number of interaction events (for a detailed discussion, see the Supporting Information). 
 Our framework allows us to model more general situations by explicitly choosing $P_{l|k}$. 

In order to test our analytical model, we considered epidemiological dynamics taking place on artificial networks on which we release the constraints of linearity found in  'classical' networks. To create these networks, we used combinations of Poisson and  power law distributions for the number of contacts $k$ and interaction events per time $l$.  This allows us to introduce arbitrary combinations of homogeneous or heterogeneous behaviour in the way contacts are made and in the number of interaction events established, that may be either independent or dependent (as in the linear case).

We generated four types of networks corresponding to the combinations of homogeneous and heterogeneous behaviour in the number of contacts $k$ and the number of interaction events $l$ as well as the corresponding networks with a linear dependency between $k$ and $l$. We studied the spread of an infectious agent on these artificial networks using the analytical approach by plugging the corresponding joint probability generating functions into equations \ref{eq:s2}-\ref{eq:s3} (further details concerning the PGFs used for each network are given in the Materials and Methods section). In addition, we ran numerical simulations. For these simulations, networks were generated by assigning each host a number $k$ of `half-contacts'  (stubs) and  $l$ interaction events  per time interval drawn from the distribution $P_{kl}$. Each host then distributed his/her  interaction events equi-probably at random (i.e.multinomially) among his/her $k$ contacts. $P_{kl}$ was chosen to suit $\langle l \rangle =2\langle k\rangle$ corresponding to a timescale in which a host has on average 2 interaction events per contact. 

 The next step in the generation of a configuration type network consisted in randomly matching half-contacts (stubs) respecting their assigned number of  interaction events per contact. 
This may introduce topological constraints on the network that can lead to network segregation and assortative effects that are not seen in the analytical approach \emph{per se} due to its node-centric view (see Supporting Information).

\begin{figure*}[b!]
\includegraphics[angle=0, height = 11cm]{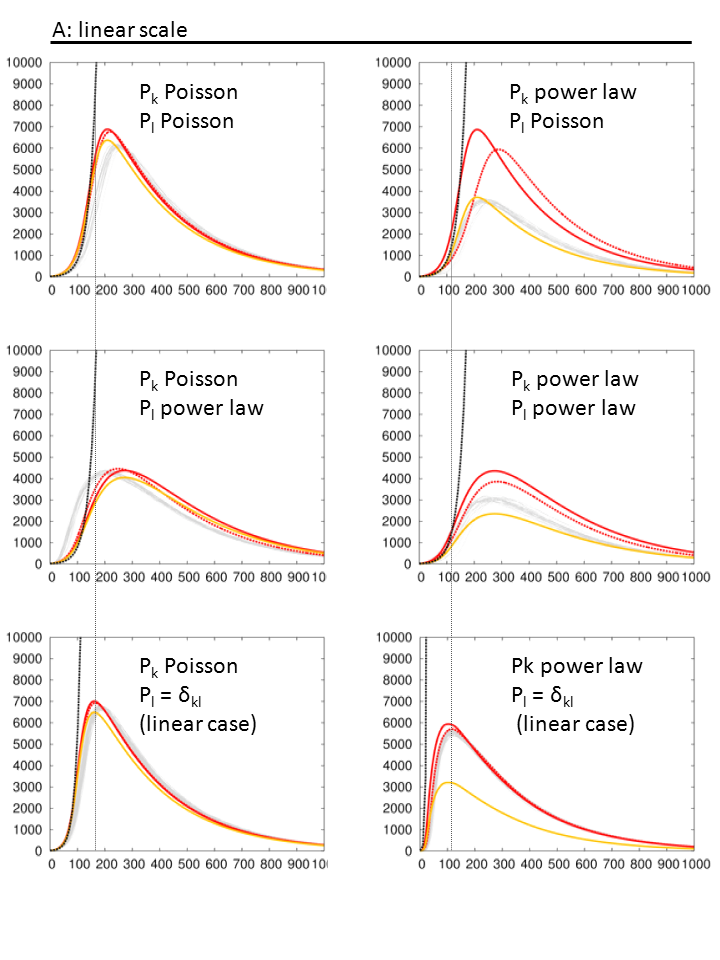}
\includegraphics[angle=0, height = 11cm]{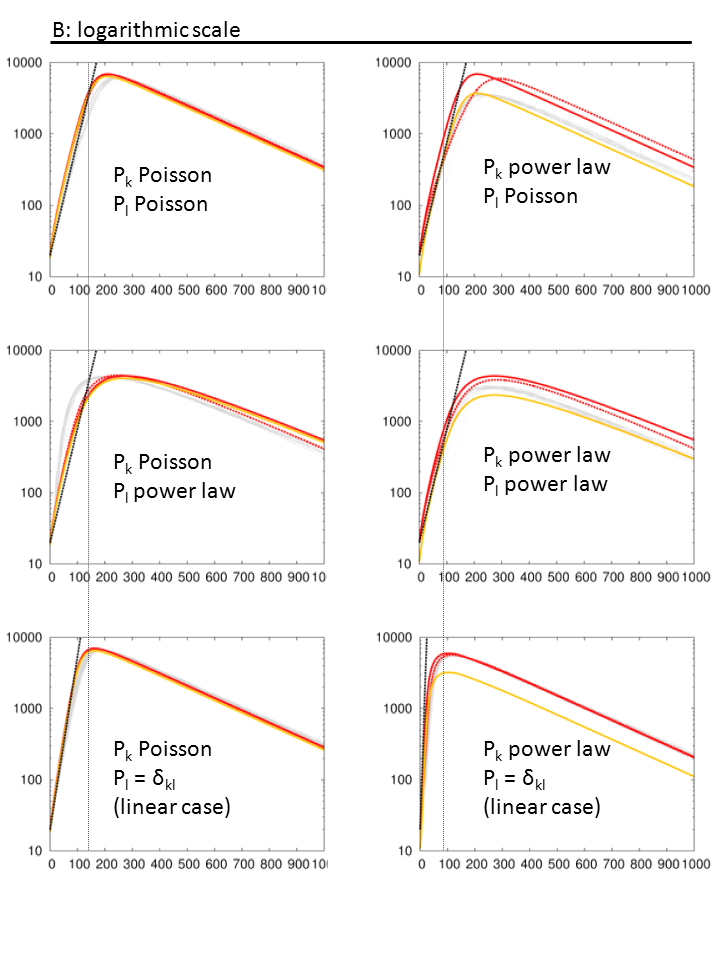}
\caption{ Dynamics of the number of infected hosts ($\scriptstyle I$) during epidemic spreading on different types of networks. The distributions in the number of contacts ($\scriptstyle k$) and interaction events per time ($\scriptstyle l$) are either homogeneous (Poisson) or heterogeneous (power law) For the number of interaction events, we also show the linear case in which { $\scriptstyle l$} is strictly proportional to the number of  contacts $\scriptstyle k$. $\scriptstyle k$ and $\scriptstyle l$ are drawn from joint distributions $\scriptstyle P_{kl}$ with $\scriptstyle \langle l\rangle = 2\langle k\rangle$ (except for the analytical $\scriptstyle P_{kl}$ model's linear case where $\scriptstyle l = k$ being compensated by a double transmission rate.). The figures show the outcome of the simulation runs (grey, dotted lines), of the the numerical solution of the analytical model with $\scriptstyle P_{kl}$ (red, solid line) and $\scriptstyle \bar{P}_{kl}$ (red, dashed line). In addition, we show the epidemic prevalence when excluding individuals with only one contact ($\scriptstyle k=1$) which is relevant for epidemics on networks with heterogeneous numbers of contacts including many individuals with $\scriptstyle k=1$ in combination with a (nearly) constant number of interaction events, as realised through a Poisson distribution (orange line, cf. specifically $P_k$ power law, $P_l$ Poisson). Parameters chosen correspond to $\scriptstyle \langle k\rangle = 4$ (Poisson case: $\scriptstyle \langle k \rangle = 4$, $\scriptstyle \langle l\rangle = 8$,  power law case: $\scriptstyle \lambda_k = 1.4$, $\scriptstyle \lambda_l=0.89$, $\scriptstyle \kappa_k = \kappa_l = 22$). Epidemiological parameters are $\scriptstyle \beta=0.01$ ($\scriptstyle 0.02$ for the analytical $\scriptstyle P_{kl}$ model's linear case), $\scriptstyle \gamma = 0.004$ in arbitrary units and $\scriptstyle I(0)=20$. Early epidemic expansion is further captured by exponential growth with $\scriptstyle I(0)e^{r_0 t}$ (black, dashed line).   \label{fig1}}
\end{figure*}

If the number of  interaction events per contact is large ($\langle l\rangle/\langle k\rangle\gg1$), hosts nearly equi-distribute their  interaction events among their contacts due to convergence under the law of large numbers. Exact matching of the host's number of interaction events per contact and time segregates the network into subnetworks according to this number (weight assortativity). 
A more realistic network is obtained by decreasing $\langle l \rangle$ and thereby introducing some variability in the number of interaction events among a host's contacts which simulates some flexibility in assigning interaction events at short time scales. 
In addition, we also introduced some tolerance in `negotiating' the number of interaction events per contact, which leads to a slightly modified empirical joint probability distribution (denoted $\bar{P}_{kl}$ with PGF $\bar{G}(x,y)$) which is discussed in detail in the Materials and Methods. 
Weight assortativity in terms of $l/k$ leads to a network segregation that constrains epidemics but it might at the same speed up the initial expansion of an epidemic as compared to the prediction of the analytical approach. This situation arises in networks with a (nearly) constant number of contacts $k$ per individual among which  a heterogeneous number of interaction events $l$ is distributed leading to an early expansion among mostly highly active individuals. 
Alternatively, the network can also segregate with respect to the number of contacts an individual holds (contact or degree assortativity). An extreme case can be observed when a (nearly) constant number of interaction events has to be distributed among a heterogeneous number of contacts. This leads to a (near) isolation of individuals with single contacts from the epidemic process (for details cf. Supporting Information).

To validate the model, we compared the  epidemic prevalence ($I$) from repeated simulation runs  with the results derived from the analytical approach using the probability generating functions corresponding to $P_{kl}$ and $\bar{P}_{kl}$, $G(x,y)$ and $\bar{G}(x,y)$ (Note that $P_{kl}(0)=P_{Skl}(0)$, $\bar{P}_{kl}(0)=\bar{P}_{Skl}(0)$, $G(x,y,0) = G_S(x,y,0)$ and $\bar{G}(x,y,0)=\bar{G}_S(x,y,0) $.)
The epidemiological dynamics are summarized in Fig. \ref{fig1} through a comparison of the simulation results with solutions of the analytical approach for $P_{kl}$ ($G(x,y)$), $\bar{P}_{kl}$ ($\bar{G}(x,y)$) and an approximation (applied to $P_{kl}$) excluding individuals with one contact as relevant for networks with heterogeneous number of contacts and (nearly) constant number of interaction events per individual (contact assortativity). The analytical approach represents the simulation results well if the constraints of the specific networks topology are properly taken into account. A particularly interesting observation is that epidemics in networks with heterogeneous contact behaviour slow in independently weighted networks ($P_{kl}=P_kP_l$) as compared to the case of 'classical, linar' networks in which the number of interaction events scales with the number of contacts for each individual.

\section{Capturing epidemic characteristics ($\scriptstyle r_0$ and $\scriptstyle R_0$)}

\begin{table*}[b!]
\caption{$r_0$ vs. $R_0$}\label{R0r0}
\begin{tabular*}{\hsize}
{p{3cm}p{7cm}p{5cm}}
 & Early epidemic growth rate $r_0$& Basic reproductive ratio $R_0$\\
\hline
Epidemic expansion from randomly picked index case, (mean field approximation)
$M_{SI}\approx I \langle k\rangle $&  \begin{eqnarray} \dot{I}&\approx&(\beta \, \langle l\rangle-\gamma) I\nonumber\\ r_0^{MF}&=& \beta \, \langle l\rangle -\gamma \nonumber \end{eqnarray}& -\\
\hline
Early epidemic expansion, structure set by epidemic, $M_{SI}\approx I \left(\frac{\langle kl\rangle }{\langle l\rangle }-2\right) $ & 
\begin{eqnarray} \dot{I}&\approx & \left( \frac{\langle kl\rangle -2\langle l\rangle }{\langle k\rangle } \, \beta -\gamma\right) I \nonumber \\
&=& \left(\left(\frac{\langle kl\rangle }{\langle l\rangle }-2\right) \frac{\langle l\rangle}{\langle k\rangle} \, \beta-\gamma\right) I \nonumber \\
r_0&=&  \left(\frac{\langle kl\rangle }{\langle l\rangle }-2\right) \frac{\langle l\rangle}{\langle k\rangle} \, \beta -\gamma \nonumber\end{eqnarray}  
& 
\begin{eqnarray}
R_0&=&\frac{\langle \frac{l}{k}\rangle \, \beta}{\langle \frac{l}{k}\rangle \, \beta +\gamma}\left(\frac{\langle kl\rangle }{\langle l\rangle}-1\right)\nonumber
\end{eqnarray}
\\
\hline
\end{tabular*}
\end{table*}
There are different ways to assess the initial propagation of an infectious agent in an otherwise fully susceptible population. One possibility is to estimate the initial exponential growth rate $r_0$ in the number of infected individuals. Another possibility consists in estimating the number of secondary cases created by a newly infected  host in a fully susceptible population,  which is classically referred to as the basic reproductive ratio $R_0$ \cite{AndersonMay1991}.  

Furthermore, subtle effects arise depending on whether we take the neighbourhood of  a random individual as a reference or the neighbourhood of a `typically' infected individual. The first case (using a random individual as reference) corresponds to what is usually referred to as a `mean field approximation' and describes the very first infection events. The second case (using a `typical' infected individual) is more appropriate to capture the next stages of early epidemic expansion because it accounts for the fact that spatial structure has been sensed or set by the epidemic process and is considered for the derivation of $R_0$.   

The expressions for $r_0$ and $R_0$ for the SIR model are shown in Tab.~\ref{R0r0} and are derived in detail in the sections on Materials and Methods and the Supporting Information, respectively. Note that these do not involve any approximation beyond those implied in the model's assumptions, i.e.~they are exact within the model framework. The derivation for $R_0$ in the case  where $\gamma>0$ is the only one that  required some further approximations to obtain an explicit formula (cf. Supporting Information). Fig.~\ref{fig1} shows that the exponential growth rate $r_0 = \left(\frac{\langle kl\rangle }{\langle l\rangle }-2\right) \frac{\langle l\rangle}{\langle k\rangle} \, \beta -\gamma$ calculated for the early epidemic expansion corresponds well to the simulation and analytical results. The expressions for $r_0$ and $R_0$ scale with the second moment $\langle kl \rangle$ of the joint probability distribution $P_{kl}$ quantifying how correlations between the number of contacs $k$ and interaction events $l$ an individual maintains affect epidemic spreading. 

\section{Real world scenarios - application to epidemiological data}

\begin{figure}[h!]
\includegraphics[angle=0, height = 12cm]{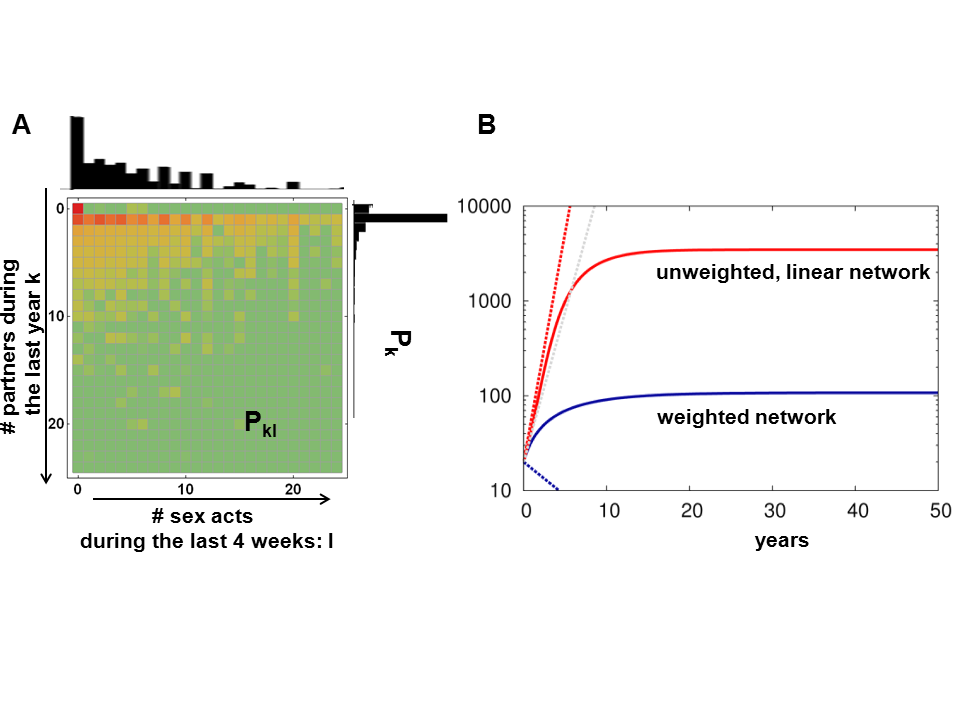}
\caption{A) Characteristics of the sexual contact network inferred from the NATSAL contact tracing study and B) Dynamics of SI epidemics spreading on sexual contact networks. The network of heterosexual contacts shows a heterogeneous joint probability distribution $\scriptstyle P_{kl}$ for an individual to have $\scriptstyle k$ contacts and $\scriptstyle  l$ sex acts (interaction events) as derived from NATSAL data \protect\cite{JohnsonEtal2001}, which holds analogously for the marginal distributions $\scriptstyle P_k$ and $\scriptstyle P_l$ (panel A). On Panel B, SI epidemics with transmission rate per sex act of $\scriptstyle  \beta = 0.01$ along a weighted network of the sexual contacts as defined by $\scriptstyle P_{kl}$ (blue, dashed line corresponds to early exponential approximation $\scriptstyle I(0)e^{rt}$) and along a classical network with degree distribution $\scriptstyle P_k$ (red, dashed line corresponds to early exponential approximation $\scriptstyle I(0)e^{r_0t}$). Note that the epidemic is self-maintained on the classical network ($\scriptstyle r_0>0$) while it is not on the weighted network ($\scriptstyle r_0<0$). In the latter case, only a minor epidemic spreads around randomly infected hosts as $\scriptstyle r_0^{MF}$ exceeds zero for these initial nodes (grey line, dashed, mean field-approximation  $\scriptstyle r_0^{MF}$).  }\label{NATSAL}
\end{figure}

The knowledge of transmission networks along which an infectious agent can spread within a host population is of great importance to public health. These networks might be  hard to define for air-borne infections because they are very dynamic \cite{EdmundsKafatos2006, WallingaTeunis2006} but easier to define for sexually transmitted infections (STI) because they are rather static. Such sexual contact networks have been surveyed in many studies covering homosexual as well as heterosexual  populations and different societal contexts \cite{SchneebergerEtal2004,JohnsonEtal2001,HamiltonEtal2008,Carael2009} to understand and confine the spread of  STI. The National Survey of Sexual Attitudes and Lifestyles (NATSAL \cite{JohnsonEtal2001}) provides very detailed data on the situation in the United Kingdom including distributions in the number of sexual partners $k$ and sex acts (interaction events $l$) a person has within certain time frames.  As shown in Fig.~\ref{NATSAL}A, both the number of partners (contacts $k$) and sex acts (interaction events) $l$ an individual has are heterogeneously distributed. However, their joint distribution $P_{kl}$ does not show a linear behaviour, meaning that the number of sex acts $l$ does not scale linearly with the number of partners $k$ an individual has. Still, as pointed out in \cite{Moslonka-LefebvreEtal2012},  many epidemic modelling studies on unweighted networks rely on this assumption. The linearity assumption in combination with the broad distribution found in the number of sexual contacts and sex acts results in predictions of very fast early epidemic expansion and an epidemic threshold that is potentially vanishing in the limit of infinite network size \cite{Durrett2007, MayAnderson1987,Moreno2002,BogunaVespignani2003,BansalEtal2007,VolzMeyers2009} as can be seen from equation \ref{R0classical}. 

In our `Validation' section we have shown that, regarding the number of interaction events, a deviation from the linear behaviour results in a decrease in epidemic expansion and peak prevalence, especially for transmission networks that are characterized by a heterogeneous distribution in the number of contacts per individual $k$ and regardless of whether the distribution of interaction events/sex acts is homogeneous (Poisson) or heterogeneous (power law). This behaviour is also reflected in Fig.~\ref{NATSAL}, which shows the epidemic expansion of a susceptible-infected epidemic with transmission rate $\beta=0.01$ per sex act in two network scenarios. The first scenario (red curve) builds on a linear contact network only defined by the contact distribution $P_k$, in which the number of sex acts grows linearly with the number of contacts respecting the average number of sex acts ($\langle l \rangle = 5.59$ during 4 weeks). The second scenario (blue curve) takes the joint probability distribution $P_{kl}$ into account and evaluates the epidemic on the resulting weighted network. Only by looking at the exponential growth rates of the epidemics, which are $r_0=1.111$ per year and $r_0=-0.1625$ per year, respectively, one can conclude that the former scenario supports epidemic expansion, while the latter does not beyond a few infections caused by the initially randomly infected individuals ($r_0^{MF}=\beta\langle l\rangle = 0.1679$ per year in the mean field approximation, grey dashed line). Although the survey data shown in Fig.~\ref{NATSAL}A  only provides us with a rough picture of the real transmission network and although the number of partners during one year can only approximately be considered as being concurrent, the data are sufficient to confirm a remarkable reduction in the epidemic potential when shifting from a classical unweighted transmission network towards a more realistic weighted transmission network. This finding is in particular consistent with an earlier simulation study on epidemic spreading along a network of homosexual contacts \cite{Moslonka-LefebvreEtal2012}.

\section{Discussion}
Network theory and specifically network epidemic models have broadened our understanding of the spread of infectious agents - or other entities such as information, money, travellers or goods - in complex settings. In their simplest form, they do not consider that contacts may show variability in their transmission capacity. But the probability of disease transmission along a contact strongly depends on the intensity of the contact, tansportation links vary in their throughput and information may not be shared equally among all possible channels. Although earlier studies have shown that this weighting of contacts has non-negligible impact on the spreading dynamics \cite{Moslonka-LefebvreEtal2012} modelling of of epidemics on weighted networks largely focuses on simulation studies \cite{EamesEtal2009,ChuGuanEtal2009,Moslonka-LefebvreEtal2012}, mean field approximations \cite{ChuEtal2009,YangEtal2012} or discrete time dynamics \cite{BrittonEtal2011, Deijfen2011} providing explicit expressions for epidemic characteristics such as the basic reproductive ratio $R_0$ only in special cases. 

We extend these findings by providing a framework based on partial differential equations that allows to model continuous time SIR epidemic dynamics for very general weighted networks defined through the joint probability distribution for an individual to have $k$ contacts and $l$ interaction events. From this we are able to derive the full epidemic dynamics in terms of the number of susceptible, infected and recovered individuals over time as well as explicit expressions for the basic reproductive ratio $R_0$ and the exponent of early epidemic expansion $r_0$. The application of our method to epidemics on artificial and empirically motivated networks corresponds well to simulation results on these networks. Moreover, it reveals the impact of assortative effects introduced by contact weighting on epidemic dynamics which will need closer attention in future research.

\section{Materials and Methods}
\subsection{Analytical results and their validation}
The set of differential equations describing the epidemic process is derived by careful bookkeeping of the links along which an infectious agent spreads and a detailed derivation is provided in the Supporting Information. The analytical approximation assumes that an individual distributes his/her  $\scriptstyle l$ interaction events multinomially among his/her $\scriptstyle k$ contacts and is infected at a rate proportional to his/her average number of interaction events with $\scriptstyle i$ infected contacts. This averaging implies the choice of a time scale on which $\scriptstyle \langle l \rangle > \langle k \rangle$. As the analytical approach is node-centric it does however not consider the constraints for half-contacts to match half-contacts with similar weight leading to an unrealistic network segregation in some artificial networks for $\scriptstyle \langle l \rangle > \langle k \rangle$. This network segregation changes epidemic dynamics in these networks which is not seen in the analytical approach \emph{per se} but can be considered through corrections in the analytical approach.

To validate the analytical approach, we look at combinations of Poisson and power law distributions for the number of contacts $\scriptstyle k$ and interaction events per time $\scriptstyle l$. As we neglect isolated hosts, we look at the Poisson distribution  $\scriptstyle P_n=\frac{1-\delta_{n0}}{1-e^{-\langle n \rangle}}\frac{\langle n\rangle^n}{n!}e^{-\langle n \rangle }$ with support for $\scriptstyle n>0$ and probability generating function $\scriptstyle g(x)=\frac{e^{\langle n\rangle(x-1)}-e^{-\langle n\rangle}}{1-e^{-\langle n\rangle}}$ and power law distributions with exponent $\scriptstyle \lambda$ and cut-off $\scriptstyle \kappa$, $\scriptstyle P_n=\frac{n^{-\lambda}e^{-\frac{n}{\kappa}}}{\text{Li}_{\lambda}\left(e^{-\frac{1}{\kappa}}\right)}$ and $\scriptstyle g(x)=\frac{\text{Li}_{\lambda}\left(xe^{-\frac{1}{\kappa}}\right)}{\text{Li}_{\lambda}\left(e^{-\frac{1}{\kappa}}\right)}$ (normalisation through the Polylogarithm $\text{Li}_\lambda$) for homogeneous and heterogeneous behaviour, respectively.
If the joint probability distribution $P_{kl}$ is given by the product $P_{kl}=P_kP_l$ of independent distributions with PGFs $g_1(x)$ and $g_2(y)$ their joint PGF $G(x,y)$ is also given by the product $G(x,y)=g_1(x)g_2(y)$. In the linear case with $P_{kl}=P_k\delta_{kl}$ the PGF $G(x,y)$ is given by $G(x,y)=g_1(xy)$.

\subsection{Derivation of $r_0$ and $R_0$} The derivation of $r_0$ is based on the observation that the rate of epidemic expansion as described by equations (\ref{S}-\ref{R}) is proportional to the number of links/contacts between susceptible and infected individuals $M_{SI}$, i.e. $p_{SI} \, S\, \langle l\rangle_S = \frac{M_{SI}}{M_S} \,  S \, \langle l\rangle_S  = M_{SI}  \, \frac{\langle l\rangle_S}{\langle k\rangle_S}\approx M_{SI} \frac{\langle l\rangle}{\langle k\rangle}$, with $M_S=S  \, G_S^{(1,0)}(1,1,t)=S \langle k \rangle_S$ being the number of contacts of susceptible individuals. 

In the mean field approximation, $M_{SI}$ is approximated by the number of infected individuals ($I$) and the average number of contacts per individual found originally in the total population ($\langle k\rangle =G^{(1,0)}(1,1,0)$). As soon as the epidemic is set, $M_{SI}$ is evaluated from the product of $I$ and a slightly more sophisticated estimate of the number of contacts of infected hosts: each infected node contributes to $M_{SI}$ by the average excess degree of a recently infected node $\frac{G^{(1,1)}(1,1,t)}{G^{(0,1)}(1,1,t)}-1=\frac{\langle kl\rangle}{\langle k\rangle}-1$ (for details, see the Supporting Information). This means discounting the contact the infection has spread from and assuming that all `new' contacts are still susceptible in the early phase of an epidemic. Considering that the contact along which the epidemic spreads further also needs to be discounted results in $M_{SI}\approx I\left(\frac{G^{(1,1)}(1,1,t)}{G^{(0,1)}(1,1,t)}-2\right)= I\left(\frac{\langle kl\rangle}{\langle k\rangle}-2\right)$.

All together  we have,
\begin{eqnarray}
\dot{I} \!\!\!\!\!\!\!\!\!\!\!\!\!&=&\!\!\!\!\!\! rp_{SI} \langle l\rangle_S S  - \gamma I\nonumber\\
&=& \!\!\!\!\!\!rM_{SI}\frac{\langle l\rangle_S}{\langle k\rangle_S} -\gamma I\nonumber\\
&\substack{M_{SI}\approx I \langle k\rangle \\ \approx}&  \!\!\!\!\!\!(r\langle l \rangle - \gamma)I\label{mf}\\
&\substack{M_{SI}\approx I \left(\frac{\langle kl\rangle}{\langle k\rangle}-2\right) \\ \approx}& \!\!\!\!\! \left(\left(\frac{\langle kl\rangle -2\langle l\rangle}{\langle k\rangle }\right)r-\gamma\right) I.\label{2}
\end{eqnarray}
Approximation (\ref{mf}) corresponds a ``mean field approximation'' representing the neighbourhood of a randomly picked node, i.e. not a node picked according to its number of  interaction events per time interval. Approximation (\ref{2}) considers that an infected individual has been picked with a probability proportional to its number of  interaction events per time interval. The doubling time $t_D$ can be derived from $r_0$ as $t_D=\ln (2)/r_0$.

The basic reproductive ratio $R_0$ is the average number of secondary infections of a typical infected host in a fully susceptible population. As for SIR models on classical random networks, it is derived by first evaluating the distribution of excess contacts of a typically infected host, i.e.~the probability for a node chosen proportional to its number of  interaction events per time ($l$) to have $k$ excess contacts. This probability is $Q_{kl}=\frac{lP_{(k+1)l}}{G^{(0,1)}(1,1)}$. 
$R_0$ is evaluated in the Supporting Information as the number of infections that spread along these excess contacts before recovery of the typically infected host.

\subsection{Simulations} 
Networks for simulation were generated by first generating 10000 nodes with $k$ half-contacts (stubs) and $l$ interaction events as drawn from the probability distribution $P_{kl}$. Each nodes $l$ interaction events are then distributed multinomially among its $k$ stubs. Stubs are randomly matched with matches being rejected if the stubs' weights differ by more than one interaction event or more than $10\%$. If the stubs with non-identical weights are matched, the contact is assigned the mean weight randomly rounded to the next integer. This results in the empirical distribution $\bar{P}_{kl}$.

We use the Gillespie direct algorithm \cite{KeelingRohani2008Book} to run stochastic SIR epidemics on continuous time. For each susceptible node $i$, transmission occurs at rate $\sum\limits_{j \in I} {\beta {W_{j,i}}}$ where $\beta$ is the probability of transmission per sex act, $W$ is the weighted adjacency matrix listing the number of interaction events per time step between all pairs of nodes, and $j$ are the infected nodes connected to node $i$. Infected nodes recover at rate $\gamma$. For each case analyzed, 20 nodes were initially infected uniformly at random in a population of 10000 in replicate simulations and networks. 
\section*{Acknowledgments}
MML would like to thank the French Ministries in charge of Agriculture and Environment for financial support.

\clearpage
\appendix
\section{Supporting Information}

\subsection{Equations for the epidemic model on weighted networks}
We consider an epidemic of a disease that is transmitted at a rate, $\beta$, and from which infected individuals  recover at a rate $\gamma$. Susceptible individuals with $k$ contacts and $l$ interaction events per time interval are infected at a rate proportional to $\beta$, $l$ and the probability that a susceptible individual's contact is made with an infected individual $p_{SI}$. This leads to the following equations for the evolution of the number of susceptible and infected individuals with $k$ (infectious) contacts and $l$ interaction events per time
\begin{eqnarray}
\dot{S}_{kl}&= & -\beta p_{SI}lS_{kl}\label{SI:skl}\\
\dot{I}_{kl}&= & +\beta p_{SI}lS_{kl}-\gamma I_{kl}\label{SI:ikl}  \\
\dot{R}_{kl}&= & \gamma I_{kl} ,
\end{eqnarray}
for which a detailled overview of the model's notation and parameters is given in Table \ref{tab:modelnotation}.

Adding up the contributions for all $k$ and $l$ introduces the average number of  interaction events per time and susceptible individual $\langle l\rangle_S=\sum_l lP_{Skl}=\sum_l l\frac{S_{kl}}{S}$ into the equations. This average number can also be expressed in terms of probability generating function $G_S(x,y,t)=\sum_{kl}P_{Skl}(t)x^ky^l$ of the joint probability distribution to find $k$ contacts and $l$ interaction events per time among susceptible individuals $P_{Skl}$: $\langle l\rangle_S = \sum_{k,l}l\frac{S_{kl}}{S}=G_S^{(0,1)}(1,1,t)$. The $(0,1)$ exponent of $G_S$ indicates the orders of the partial derivatives with respect to the first and second argument of $G_S$ (see Table \ref{tab:modelnotation}).

Summation of $S_{kl}$ and $I_{kl}$ over $k$ and $l$ results in equations for the total number of susceptible and infected hosts: 
\begin{eqnarray}
\dot{S} &=& -\beta p_{SI} S G^{(0,1)}_S(1,1,t)   \label{SI:S}\\
\dot{I} &=& \beta p_{SI} S G^{(0,1)}_S(1,1,t)  - \gamma I\\
\dot{R} &=& \gamma I \label{SI:R}
\end{eqnarray}

To close this set of equations we also need to derive equations for $p_{SI}$, as well as for the probability generating function (PGF) $G_S(x,y,t)$. 

We begin by deriving the temporal dynamics of the probabilities for a link starting from a susceptible individual to point to a susceptible or infected individual, $p_{SS}$ and $p_{SI}$, respectively.  Following the argument in \cite{Volz2008} we write $p_{SS}={M}_{SS}/{M_S}$ and $p_{SI}={M}_{SI}/{M_S}$ to express these probabilities in terms of the number of links/contacts among susceptible and infected hosts ($M_{SS}$, $M_{SI}$) and total number of links/contacts of susceptible hosts ($M_{S}$). From this, we get:

\begin{eqnarray}
\dot{p}_{SS} &=&\frac{\dot{M}_{SS}}{M_S}-\frac{\dot{M}_{S}}{M_S}p_{SS}\\
\dot{p}_{SI} &=& \frac{\dot{M}_{SI}}{M_S}-\frac{\dot{M}_{S}}{M_S}p_{SI}
\end{eqnarray}
for which expressions are derived in the following paragraph.
From the definition of $M_S$, we can derive the following equation:
\begin{eqnarray}
\dot{M}_S&=&\sum_{k,i} k \dot{S}_{kl}
\end{eqnarray}
After substitution from equation (\ref{SI:skl}) this results in
\begin{eqnarray}
\dot{M}_S&=& -\beta p_{SI}SG^{(1,1)}_S(1,1,t)
\end{eqnarray}

We then follow the arguments made in an earlier study \cite{Volz2008}, which rely on the assumption that the number of contacts from susceptible hosts to susceptible, infected and recovered hosts is multinomially distributed with probabilities $p_{SI}$, $p_{SS}$ and $p_{SR}=1-p_{SS}-p_{SI}$. We also assume that the same applies to the number of  interaction events/sex acts per time interval. If a node with $k$ contacts has $j$ contacts with susceptible individuals and $i$ contacts with infected individuals its   interaction events with susceptible, infected and recovered individuals $n_{SS}$, $n_{SI}$ and $n_{SR}$, respectively, are distributed according to 
\begin{equation}
 \frac{l!}{n_{SS}!n_{SI}!n_{SR}!}\left(\frac{j}{k}\right)^{n_{SS}}\left(\frac{i}{k}\right)^{n_{SI}}\left(\frac{k-j-i}{k}\right)^{n_{SR}}
\end{equation}
with averages $\langle n_{SS} \rangle = \frac{jl}{k}$,$\langle n_{SI} \rangle = \frac{il}{k}$ and $\langle n_{SR} \rangle = \frac{(k-j-i)l}{k}$.  Note that $l=n_{SS}+n_{SI}+n_{SR}$.

The probability that a susceptible node with $j$, $i$ and $k-i-j$ contacts to susceptible, infected and recovered individuals, respectively, is reached from an infected node, i.e.~chosen with a probability proportional to the average number of sex acts with infected nodes ($\langle n_{SI}\rangle =\frac{li}{k}$) is then
\begin{eqnarray}
 &&\frac{P_{kl}\frac{k!}{i!j!(k-i-j)!}p_{SS}^jp_{SI}^ip_{SR}^{k-j-i}\langle n_{SI}\rangle}{p_{SI}G^{(0,1)}(1,1)}\\
&\stackrel{\langle n \rangle_{SI} = \frac{li}{k}}{=}&\frac{lP_{kl}\frac{(k-1)!}{(i-1)!(k-1-(i-1)-j)!}p_{SS}^jp_{SI}^{i-1}p_{SR}^{k-j-i}}{G^{(0,1)}(1,1)}
\end{eqnarray}
Note that the denominator can be derived using the observation that
\begin{eqnarray}
&&\sum_{k,l}P_{kl}lp_{SI}\sum_{i,j\leq k}\frac{(k-1)!}{(i-1)!(k-1-(i-1)-j)!}p_{SS}^jp_{SI}^{i-1}p_{SR}^{k-j-i}\\
&=&\sum_{k,l}P_{kl}l(p_{SS}+p_{SI}+p_{SR})^{k-1}\\
&=&p_{SI}\sum_{k,l}lP_{kl}=p_{SI}G^{(0,1)}(1,1)
\end{eqnarray}

Therefore, the probability distribution for the contacts and potential  interaction events of a node which was chosen proportional to $\langle n_{SI} \rangle = \frac{il}{k}$ is generated by
\begin{eqnarray}
 &&\frac{ \sum_{kl}lP_{kl}y^l\sum_{i,j\leq k}\frac{(k-1)!}{(i-1)!(k-1-(i-1)-j)!}(x_Sp_{SS})^j(x_{I}p_{SI})^{i-1}(x_Rp_{SR})^{k-j-i}}{G^{(0,1)}(1,1)}\\
&=& \frac{ \sum_{kl}lP_{kl}(x_Sp_{SS}+x_Ip_{SI}+x_Rp_{SR})^{k-1}}{G^{(0,1)}(1,1)}\\
&=&\frac{x_Iy}{x_Sp_{SS}+x_Ip_{SI}+x_Rp_{SR}}\frac{G^{(0,1)}(x_Sp_{SS}+x_Ip_{SI}+x_Rp_{SR},y)}{G^{(0,1)}(1,1)}
\end{eqnarray}
Choosing a node proportional to its average number of  interaction events per time ($\langle n_{SI} \rangle = il/k$)  instead of the actual number of  interaction events ($n_{SI}$) implies the assumption of a time scale at which $l\gg k$, i.e.~a case in which fluctuations around $\langle n_{SI} \rangle$ can be expected to be small.
Taking all this together, the average degrees of a susceptible node that was reached from an infected node to susceptible or infected nodes are
\begin{eqnarray}
\delta_{SI}(S)&=& \left. \frac{\partial}{\partial x_S}  \frac{x_Iy}{x_Sp_{SS}+x_Ip_{SI}+x_Rp_{SR}}\frac{G^{(0,1)}_S(x_Sp_{SS}+x_Ip_{SI}+x_Rp_{SR},y,t)}{G^{(0,1)}_S(1,1)}  \right|_{x_S=x_I=x_R=y=1} \\
&=&p_{SS}\left(\frac{G^{(1,1)}_S(1,1,t)}{G^{(0,1)}_S(1,1,t)}-1\right)\\
\delta_{SI}(I)&=& \left.  \frac{\partial}{\partial x_I}\frac{x_Iy}{x_Sp_{SS}+x_Ip_{SI}+x_Rp_{SR}}\frac{G^{(0,1)}_S(x_Sp_{SS}+x_Ip_{SI}+x_Rp_{SR},y,t)}{G^{(0,1)}_S(1,1)} \right|_{x_S=x_I=x_R=y=1}\\
& =&p_{SI}\left(\frac{G^{(1,1)}_S(1,1,t)}{G^{(0,1)}_S(1,1,t)}-1\right)+1
\end{eqnarray}
 The average number of contacts to susceptible and infected nodes need to be discounted by one for the number of contacts with infected nodes to take the contact to the source of infection into account (directly considering this in the PGF gives the same result), i.e. the total excess degree of the afore described type of node is $\frac{G^{(1,1)}_S(1,1,t)}{G^{(0,1)}_S(1,1,t)}-1$.

Bookkeeping of the changes in the numbers of links among susceptible and infected hosts due to the epidemic process results in: 

\begin{table*}[h!]
\begin{tabular}{clp{9.5cm}}
 &  & \textbf{Changes due to epidemic spread} \\
 \hline
$\dot{M}_{SI}=$ &$-\dot{S}(p_{SI}-p_{SS})\left(\frac{G^{(1,1)}_S(1,1,t)}{G^{(0,1)}_S(1,1,t)}-1\right)$ & : change in the number of susceptible nodes $\dot{S}=-\beta p_{SI} S G^{(0,1)}_S(1,1,t)$ due to the epidemic times their average excess contacts to susceptible and infected nodes \\
&$-\beta \frac{G^{(0,1)}_S(1,1,t)}{G^{(1,0)}_S(1,1,t)}M_{SI}$& : discount for link along which the infection spread\\
&$-\gamma M_{SI}$& : link loss due to disease progression\\
&&\\
\hline
$\dot{M}_{SS}=$& $\dot{S} 2 p_{SS}\left(\frac{G^{(1,1)}_S(1,1,t)}{G^{(0,1)}_S(1,1,t)}-1\right) $ &: change in the number of susceptible nodes $\dot{S}=-\beta p_{SI} S G^{(0,1)}_S(1,1,t)$ due to the epidemic times their average excess contacts to infected nodes (bi-directional) \\
\end{tabular}
\end{table*}

In summary this results in 
\begin{eqnarray}
\dot{M}_{SI}&=& \beta p_{SI}S \left(G^{(0,1)}_S(1,1,t)-{G^{(1,1)}_S(1,1,t)}\right)(p_{SI}-p_{SS})-\beta \frac{G^{(0,1)}_S(1,1,t)}{G^{(1,0)}_S(1,1,t)}M_{SI}-\gamma M_{SI}\\
\dot{M}_{SS}&=& - 2\beta p_{SS}p_{SI}S \left(G^{(1,1)}_S(1,1,t)-G^{(0,1)}_S(1,1,t)\right)
\end{eqnarray}
which finally leads to 
\begin{eqnarray}
\!\!\!\!\!\!\dot{p}_{SI} \!&=& \!\beta p_{SI}p_{SS} \frac{G^{(1,1)}_S(1,1,t)-G^{(0,1)}_S(1,1,t)}{G^{(1,0)}_S(1,1,t)}-\beta p_{SI}(1-p_{SI})\frac{G^{(0,1)}_S(1,1,t)}{G^{(1,0)}_S(1,1,t)} -\gamma p_{SI}\label{SI:psi}\\
\!\!\!\!\!\!\dot{p}_{SS} \!&=& \!-\beta p_{SI}p_{SS}\frac{G^{(1,1)}_S(1,1,t)-2G^{(0,1)}_S(1,1,t)}{G^{(1,0)}_S(1,1,t)}.\label{SI:pss}    
\end{eqnarray}
To close the set of equations we also need to derive an equation for the probability generating function (PGF) ${G}_S(x,y,t)$, which corresponds to the probability to find individuals with $k$ contacts and $l$  interaction events (e.g.~sex acts) per time interval among susceptible hosts, i.e.~$P_{Skl}$. From the definitions of the PGF and $P_{Skl}=\frac{S_{kl}}{S}$, we obtain
\begin{eqnarray}
\dot{G}_S(x,y,t)&=&\sum_{k,l}\left(\frac{\dot{S}_{k,l}}{S}-\frac{\dot{S}}{S}P_{S{kl}}\right)x^ky^i,
\end{eqnarray}
which results with equations \ref{SI:skl} and \ref{SI:S} in 
\begin{eqnarray}
\dot{G}_S(x,y,t)&=&\beta p_{SI}\left(G_S^{(0,1)}(1,1,t)G_S(x,y,t)-y G_S^{(0,1)}(x,y,t)\right).\label{SI:gs}
\end{eqnarray}
The probability generating functions $G_I(x,y,t)$ and $G_R(x,y,t)$ can be derived analogously (though they are not needed to close the system of equations).

\subsection{The basic reproductive ratio $R_0$}

The basic reproductive ratio $R_0$ of an SIR epidemics with transmission rate $\beta$ and recovery rate $\gamma$ on a classical network can be derived as \cite{Durrett2007}
\begin{eqnarray}
R_0&=&\frac{g^{(2)}(1)}{g^{(1)}(1)}\int_0^\infty\left(1-e^{-\beta t}\right)\gamma e^{-\gamma t}dt \\
&=&\frac{\beta}{\beta+\gamma}\left(\frac{\langle k^2\rangle}{\langle k\rangle}-1\right)
\end{eqnarray}
and is the product of the average excess degree of a node which was reached according to its degree and the transmissibility, i.e. the probability that an infection is spread along a link before recovery (here $\beta/(\beta+\gamma)$).
These terms do not factorize in the case where there are $l$  interaction events per node defined through the joint probability distribution $P_{kl}$. To derive $R_0$ for this case we first derive the excess degree distribution $Q_{kl}$ of a node that was reached with probability $l$ and having $k$ excess contacts and $l$  interaction events per time interval
\begin{equation}
 Q_{kl}=\frac{lP_{(k+1)l}}{G^{(0,1)}(1,1)},
\end{equation}
and get for $l$  interaction events multinomially distributed among the $k+1$ contacts (with probabilities $p_1=...=p_{k+1}=\frac{1}{k+1}$, $m_1+...+m_{k+1} =l$) 
\begin{eqnarray}
 R_0&=&\sum_{k,l} \int_0^\infty \gamma e^{-\gamma t} \sum_{m_1,...,m_{k+1}}\frac{l!}{m_1!...m_{k+1}!}p_1^{m_1}...p_{k+1}^{m_{k+1}} \sum_{j=1}^k(1-e^{-m_j \beta t})Q_{kl}dt\\
&=& \sum_{k,l} \sum_{m_1,...,m_{k+1}}\frac{l!}{m_1!...m_{k+1}!}p_1^{m_1}...p_{k+1}^{m_{k+1}}\sum_{j=1}^k\frac{m_j \beta}{m_j \beta+\gamma}Q_{kl}\label{SI:R0limit2}
\end{eqnarray}
The SI model is trivially included in the SIR network models in the limit $\gamma\to 0$ for which $R_0$ can be derived from equation (\ref{SI:R0limit2}):
\begin{eqnarray}
 R_0&=&\sum_{k,l} \sum_{m_1,...,m_{k+1}}\frac{l!}{m_1!...m_{k+1}!}p_1^{m_1}...p_{k+1}^{m_{k+1}}kQ_{kl}\\
&=&\sum_{k,l}  kl\frac{P_{k+1,l}}{G^{(0,1)}(1,1)}\\
&=& \frac{G^{(1,1)}(1,1)-G^{(0,1)}(1,1)}{G^{(0,1)}(1,1)}\\
&=& \frac{\langle kl\rangle - \langle l\rangle}{\langle l\rangle}=\frac{\langle kl\rangle }{\langle l\rangle}-1
\end{eqnarray}
As long as infected individuals stay indefinitely infected, $R_0$  is not affected by the transmission rate  and it measures whether there is a giant connected component in the network. Still, even then the effect of weighting of the contact network through the number of  interaction events $l$ is noticed.

The basic reproductive ratio $R_0$ for a SIR model on a weighted network can be approximated by
\begin{eqnarray}
 R_0&=&\sum_{k,l}  \sum_{m_1,...,m_{k+1}}\frac{l!}{m_1!...m_{k+1}!}p_1^{m_1}...p_{k+1}^{m_{k+1}} \int_0^\infty \gamma e^{-\gamma t}\sum_{j=1}^k(1-e^{-m_j \beta t})Q_{kl}dt\\
&\approx&\sum_{k,l}  \sum_{m_1,...,m_{k+1}}\frac{l!}{m_1!...m_{k+1}!}p_1^{m_1}...p_{k+1}^{m_{k+1}} \int_0^\infty \gamma e^{-\gamma t} k(1-e^{-\langle \frac{l}{k}\rangle \beta t}) l \frac{lP_{k+1l}}{G^{(0,1)}(1,1)}dt\\
&=&\sum_{k,l} kl\frac{lP_{k+1l}}{G^{(0,1)}(1,1)}  \int_0^\infty \gamma e^{-\gamma t} (1-e^{-\langle \frac{l}{k}\rangle \beta t})dt\\
&=&\frac{\langle \frac{l}{k}\rangle \beta }{\langle \frac{l}{k}\rangle \beta  +\gamma }\frac{G^{(1,1)}(1,1)-G^{(0,1)}(1,1)}{G^{(0,1)}(1,1)}.
\end{eqnarray}
Note that for the linear case with $P_{l|k}=\delta_{lk}$, we obtain $G(x,y)= \sum_{k,l}x^ky^l\bar{P}_k\delta_{kl}=\sum_k (xy)^k\bar{P}_k=\bar{G}(xy)$ and $R_0 = \frac{\beta}{\beta+\gamma}\frac{\bar{G}^{(2)}(1)}{\bar{G}^{(1)}(1)}$, which is consistent with earlier findings.

\subsection{The recovery of the classical equations in the linear case $P_{kl}=P_k\delta_{kl}$}
The set of equations for the weighted networks (equations \ref{SI:S}-\ref{SI:R},\ref{SI:psi}-\ref{SI:pss},\ref{SI:gs}) includes the case of classical network epidemic models, i.e. the linear case where $k=l$ or $P_{kl}=P_k\delta_{kl}$. Focussing on the degree distribution among susceptible hosts $P_{Sk}$ with probability generating function $g_S(x)$, the PGF of $P_{Skl}$ is given by $G_S(x,y)=g_S(xy)$. Substitution of $G_S(x,y)$ by $g_S(xy)$ results in 
\begin{eqnarray}
G_S^{(1,0)}(1,1,t) &=& g'_S(1,t)\\
G_S^{(0,1)}(1,1,t) &=& g'_S(1,t)\\
G_S^{(1,1)}(1,1,t) &=& g''_S(1,t)+g'_S(1,t)
\end{eqnarray}
and for the time evolution of $G_S(x,y)$:
\begin{equation}
 \dot{g}_S(xy,t)= \beta p_{SI}\left(g'_S(1)g_S(xy,t) -xyg'_S(xy,t)\right).
\end{equation}
Together, this leads to the set of equations for SIR dynamics on a classical configuration type network defined by the degree distribution $P_k$ \cite{Volz2008,Kamp2010}
\begin{eqnarray}
 \dot{S}&=&-\beta \, p_{SI}  \, S \, g'_S(1,t)    \\
\dot{I} &=& \beta \, p_{SI}  \, S  \, g'_S(1,t)  - \gamma \, I\\
\dot{R}&=& \gamma \, I\\
\dot{p}_{SI} \!\!\!\!&=&\!\!\!\!\beta  \,  p_{SI} \, p_{SS}  \,  \frac{g''_S(1,t)}{g'_S(1,t)}
-\beta  \,  p_{SI}  \, (1-p_{SI})  \, -\gamma \, p_{SI}\\
\dot{p}_{SS} \!\!\!\!&=&\!\!\!\!-\beta  \,  p_{SI}  \, p_{SS} \, \left( \frac{g''_S(1,t)}{g'_S(1,t)}-1  \right)\\
 \dot{g}_S(x,t)&=& \beta p_{SI}\left(g'_S(1,t)g_S(x,t) -xg'_S(x,t)\right).
\end{eqnarray}

\subsection{Network segregation and the limiting case $P_{kl}=P_k\delta_{\langle l \rangle l}$ (constant case)}
The analytical approximation assumes that an individual distributes his/her interaction events $l$ multi-nomially among his/her $k$ contacts and is infected at a rate proportional to his/her average number of potential transmission events with $i$ infected contacts. This averaging implies the choice of a time scale at which $\langle l \rangle > \langle k \rangle$. This leads to an unrealistic network segregation in some artificial networks, specifically for $\langle l \rangle \gg \langle k \rangle$, as the weights of an individual's contacts level at about $l/k$ and in case contacts are only made (broadly) respecting the contact weights.  This network segregation changes the epidemic dynamics. As the analytical approach is node-centric it does not consider the constraints on half-contacts to match half-contacts of  similar weight. In consequence, the change in epidemic dynamics due to networks segregation cannot be seen in the analytical approach.

The effect is particularly pronounced in networks with a heterogeneous degree distribution corresponding to many individuals with one contact in combination with a constant number of interaction events per individual. Degree one nodes have only one contact to assign their interaction events to which leaves their contacts on average already with twice the weight seen in individuals with two contacts. This weight separation leads to a situation that almost only allows contacts among individuals with a single contact, i.e. monogamous couples (contact assortativity). Therefore, individuals with one contact can only be infected if their partner is initially infected but not later on through the epidemic process because they are not connected to the giant component of the network. In the case of a constant number of interaction events per individual    $\langle l \rangle$ the analytical approach decouples into independent equations for all $k$ classes with $\langle l\rangle_S =\langle l\rangle$ in which epidemic prevalence grows at the same rates:
\begin{eqnarray}
\dot{S}_{k\langle l\rangle}&= & -\beta p_{SI}\langle l \rangle S_{k\langle l\rangle}\\
\dot{I}_{k\langle l \rangle }&= & +\beta p_{SI}\langle l \rangle S_{k\langle l \rangle}-\gamma I_{k\langle l \rangle}  \\
\dot{R}_{k\langle l\rangle}&= & \gamma I_{k \langle l \rangle} .
\end{eqnarray}
Due to the network segregation, epidemic prevalence is reduced in these networks at least by a factor proportional to the fraction of nodes with a single contact as compared to the standard prediction of the analytical approach as they do not participate in the epidemic process.

\end{document}